\documentclass[12pt,a4paper]{article}
\usepackage{amssymb,amsmath,amscd,epsfig}
\title{Evaporation of charged bosonic condensate in cosmology 
}
\author{
A.D. Dolgov$^{\rm a,b,c}$, F.R. Urban$^{\rm a,b}$
\\[5mm]
${\rm ^a}$ {\small\it Istituto Nazionale di Fisica Nucleare, via Scienze 39,
44100, Ferrara, Italy} \\
${\rm ^b}$ {\small\it University of Ferrara, Department of Physics, 
via Paradiso 12, 44100, Ferrara, Italy}\\
${\rm ^c}$ {\small\it ITEP, Bol. Ckeremushkinskaya 25,
117218, Moscow, Russia}\\
}
\date{}

\begin{document}

\newcommand{\be}{\begin{eqnarray}}
\newcommand{\ee}{\end{eqnarray}}
\newcommand{\bi}{\bibitem}
\newcommand{\lar}{\leftarrow}
\newcommand{\rar}{\rightarrow}
\newcommand{\lrar}{\leftrightarrow}
\newcommand{\mplq}{m_{Pl}^2}
\maketitle

\begin{abstract}

Cosmological evolution of equilibrium plasma with a condensate of U(1)-charged 
bosonic field is considered. It is shown that the evaporation of the condensate
is very much different from naive expectations, discussed in the literature, as 
well as from evaporation of non-equilibrium neutral condensate. The charged
condensate evaporates much slower than the decay of the corresponding bosons.
The evaporation rate is close to that of the cosmological expansion. The plasma
temperature, in contrast, drops much faster than usually, namely as the third 
power of the cosmological scale factor. As a result the universe becomes very 
cold and the cosmological charge asymmetry reaches a huge value.

\end{abstract}

\section{Introduction \label{s-intr}}

\paragraph{}
Bosonic condensates probably existed in the early universe and played an
important role in the cosmological history. Well known examples are the
classical real inflaton field, $\Phi$~\cite{infl},
or complex field, $\chi$, describing supersymmetric bosonic condensate,
which carries baryonic, leptonic, or some other $U(1)$-charge~\cite{affleck}. 
Evaporation of the inflaton produced particles creating primeval
plasma, while evaporation of $\chi$ could generate baryon or lepton
asymmetry of the universe. Though evaporations of a real field condensate
and a complex one share some similarities, the rates of the processes 
are very much different. In the case of the inflaton the rate of evaporation
is determined by the particle production rate and may be quite large, while
the evaporation of a charged condensate with a large charge asymmetry
is much slower and is determined by the universe expansion rate 
$H=\dot a/a$ during most of its history.
The impact of a large charge asymmetry on the process of evaporation has
been considered in ref.~\cite{ad-dk} (see also~\cite{dksy}). A proper account
of thermal equilibrium with a large charge asymmetry
strongly changes results of refs.~\cite{chung}-\cite{alla}, where 
evaporation of bosonic condensate was considered. These
works are applicable to the case of evaporation of uncharged condensate
(e.g. inflaton) but evaporation of a charged condensate proceeds in a
much different way, determined by thermal equilibrium 
with a large chemical potential and not just
by the rate of particle production into plasma.  

Here we elaborate the approach of ref.~\cite{ad-dk} and present more
accurate and detailed calculations. It is found in particular that the
plasma temperature in the presence of condensate drops very fast, as 
$T\sim 1/a^3$, where $a(t)$ is the cosmological scale factor. After the
condensate disappears the cooling returns to the standard relativistic
law, $T\sim 1/a$. We calculated the magnitude
of the cosmological charge asymmetry produced as a result of the 
evaporation of the condensate and found that it might be much larger than
unity because of the fast cooling of the plasma.

\section{Pre-equilibrium evolution
\label{s-pre-eq}}

\paragraph{}
We consider a charged classical scalar field $\chi$ at the final stage
of its evolution, when its amplitude is small enough and 
the potential is dominated by the quadratic term
\be
U(\chi) = m^2 |\chi|^2,
\label{U-of-chi}
\ee
while quartic or higher terms can be neglected. The equation of motion
for homogeneous $\chi$ has the form:
\be
\ddot \chi +3H\dot\chi + m^2 \chi = 0,
\label{eqn-chi}
\ee
where $H=\dot a/a$ is the Hubble parameter and $a(t)$ is the cosmological
scale factor. 

When the mass $m$ of $\chi$ becomes larger than $H$ the solution to
equation (\ref{eqn-chi}) becomes oscillating:
\be
\chi (t) = \chi_1 \cos mt + \chi_2 \sin mt
\label{sol-chi}
\ee
where the coefficients $\chi_{1,2}$ are slowly varying functions of time.
In particular, at the matter dominated (MD) regime they behave as
$\chi_{1,2} \sim a^{-3/2}$. Such a regime would be realized if cosmic
energy density is dominated by the field $\chi$ itself.
The energy density of the latter is equal to
\be
\rho_{\chi} = m^2 \left( |\chi_1|^2 + |\chi_1|^2 \right)
\label{rho-chi-cl}
\ee
The charge density of $\chi$ is expressed through the functions
$\chi_{1,2}$ as:
\be
Q_{\chi} = m \left( |\chi_1|^2 - |\chi_1|^2 \right)
\label{Q-chi-cl}
\ee
It is clear that $\chi_1$ and $\chi_2$ correspond respectively to 
the contribution of particles and antiparticles. 
For what follows we will introduce a dimensionless
parameter $\kappa$ which describes
the relative magnitude of the charge density:
\be 
\kappa = \frac{mQ_\chi}{\rho_\chi} = 
\frac{ |\chi_1|^2 - |\chi_1|^2}{ |\chi_1|^2 + |\chi_1|^2}
\label{kappa}
\ee
Evidently, at this stage $|\kappa| \leq 1$. 

The oscillating field $\chi$ starts to produce lighter particles to which
it is coupled (in many cases this can be understood as a decay of 
massive $\chi$-bosons at rest into light particles). This process is
characterized by the decay width $\Gamma$. Here we will not consider the
kinetics of the decay and do not need to specify the magnitude of
$\Gamma$. It is assumed usually, but often {\it incorrectly},
that the condensate decays with the 
rate $\Gamma$ and to the moment when $H<\Gamma$ it practically disappears
transforming its energy and charge into those of light particles. The
evolution of the energy densities of the condensate and relativistic plasma
in this case are described by the equations:
\be
\dot \rho_c = -3H \rho_c -\Gamma \rho_c \nonumber \\
\dot \rho_r = -4H \rho_r +\Gamma \rho_c 
\label{drho-c}
\ee
and indeed $\rho_c$ seems to disappear when the time is large enough,
$t\sim 1/H > 1/\Gamma$.
However, in the case when the charge density of the condensate is large, 
$\kappa \sim 1$, a large chemical potential prevents from fast evaporation 
of a charged condensate~\cite{ad-dk}. To describe this phenomenon properly 
one needs to include the contribution of inverse reactions into
eq.~(\ref{drho-c}). The problem is rather complicated and we postpone it
for the future. However, one can still use eqs. (\ref{drho-c}) at the 
initial stage of the process when the amount of relativistic particles
is small and inverse reactions can be neglected. 

Here we consider simpler but practically 
interesting case when a partial decay of the condensate creates thermally
equilibrium plasma which cools down by the universe expansion. We will
show that the evolution of such plasma and, in particular, the decay rate
of the condensate is determined by the cosmological expansion rate, $H$,
which is much slower than the decay rate $\Gamma$. The rate of the decay 
of a charged scalar field into vacuum was considered in ref.~\cite{pawl}.
Parameter $\kappa$ introduced above (\ref{kappa}) is convenient for the 
description of such equilibrium system. However, the value of $\kappa$
at the equilibrium stage should be larger than its initial value. Indeed 
the density of a conserved charge
drops down as $1/a^3$, while the energy density of mixed
non-relativistic and relativistic matter drops somewhat faster. As a result
$\kappa$ rises and can even become larger than 1. To estimate its variation
we use eqs. (\ref{drho-c}). They can be analytically solved giving:
\be 
\frac{\kappa}{\kappa_{in}} =
\left[ e^{-\eta} + \int_0^\eta d\eta' e^{-\eta'}\,
\frac{a\left(\eta'+\eta_{in}\right)}
{a\left(\eta'+\eta_{in}\right)}\,\right]^{-1},
\label{kappa-evol}
\ee
where $\eta=\Gamma (t-t_{in})$ and $\eta_{in}=\Gamma_{in} t$. 
If the universe expansion
is dominated by the non-relativistic condensate, the scale factor behaves
as $a(t) = (t/t_{in})^{2/3}$ and the integral can be easily evaluated. For 
small $\eta$: ${\kappa}/{\kappa_{in}} \approx 1 +0.4 \eta $, while 
asymptotically $ {\kappa}/\kappa_{in}\sim a $, the latter is evident. 
However, as we have already mentioned above,
the regime (\ref{kappa-evol}) remains true only till the inverse
reactions can be neglected i.e. till equilibrium is established.

Now we will show that for a large magnitude of the condensed field, 
when the particle number density in the condensate, $n_c = C(t)m^3$  
is large, i.e. $C\gg 1$,  
thermal equilibrium is quickly established when $\eta \ll 1$ and thus 
at that moment $|\kappa|$ cannot noticeably exceed unity. 
We assume that initially the cosmological matter
was dominated by the condensate of the
field $\chi$ and the contribution from light particles
(quarks, leptons and photons) can be neglected. The kinetic equation 
governing the production of light particles reads
\be
\dot n_0 + 3H n_0 = \Gamma C m^3
\label{dot-n}
\ee
where $n_0$ is the number density of the produced massless particles. The 
number density of $\chi$-particles in the condensate initially drops as 
$C(t)=C_{in}/a^3$ and the solution to equation (\ref{dot-n}) is easily found:
\be
n_0 (t) = \Gamma m^3 C_{in} \, \frac{t-t_{in}}{a^3(t)}
\label{n-0}
\ee
Thermal equilibrium would be established when $n_0 (t)$ becomes of the order of
its equilibrium magnitude, $n_0 (t) \geq n_0^{(eq)}$. To this end the following
condition is necessary:
\be
\frac{2}{3}\,\sqrt{\frac{3C}{8\pi}}\,\frac{m^3}{n_0^{(eq)}}\,
\frac{\Gamma m_{Pl}}{m^2} > 4,
\label{equil-cond}
\ee
which is evidently satisfied for a large region of the parameter values.
The mass of $\chi$-field is, of course, much smaller than the Planck mass.
It normally can be in the range $10^3-10^{10}$ GeV. 
The time during which the equilibrium is established is quite short, 
\be
\Gamma (t_{eq} -t_0) = \frac{n_0^{(eq)}}{m^3 C} \ll 1. 
\label{gamma-t-eq}
\ee

On more comment may be in order here. From kinetic equations it follows that
the number density of relativistic particles and $\chi$-particles should decrease
as $1/a^3$, if the system is very close to equilibrium. However, as we see in what
follows, the number densities of quarks or leptons remains constant during rather
large period, because they are proportional to cube of their chemical 
potential, eq. (\ref{L}), which remains constant for large asymmetry, $\mu= m$,
where $\mu$ and $m$ are respectively chemical potential and
the mass of $\chi$ (see the following section).
Similar behavior is true for their energy densities, which naively should
drop down but remain constant, see eq. (\ref{rho-rel}).
This kind of evolution is created by deviations from equilibrium
which are proportional to $H/\Gamma$ and small,
but they give corrections of the order of unity to the collision integrals 
because of a large magnitude of the latter.

\section{Thermal equilibrium \label{s-therm-equil}}

\paragraph{}
We consider the case when $\chi$ decays into the following channel:
\be
\chi \rar 3q+l
\label{chi-decay}
\ee
where $q$ and $l$ are quarks and leptons respectively. Analogous process
exists for antiparticles. Leptonic, $Q_L$, and 
baryonic, $Q_B$, charges are assumed to be conserved in this decay. Thus 
the accumulated charge of the condensate is transformed into charges of 
light particles. This is the essence of baryogenesis scenario of 
ref.~\cite{affleck}.

Particle distribution functions in thermal equilibrium are described
by two parameters, their common temperature $T$ and chemical potentials
$\mu_j$, which could change with time. As is well known, the equilibrium 
distribution of fermions is:
\be
f_{f} = \frac{1}{\exp\left(\frac{E_f-\mu_f}{T}\right)+1}
\label{f-f}
\ee
and that of bosons is
\be
f_{b} = \frac{1}{\exp\left(\frac{E_b-\mu_b}{T}\right)-1}
\label{f-b}
\ee
where $E$ are energies of the corresponding particles.
Fermionic chemical potentials may have arbitrary values, while chemical
potential of bosons is restricted from above by the value of the boson 
mass, $\mu_b <m$, otherwise the distribution function $f_b$ would not be
positive. If charge asymmetry of bosons is so big that even with maximum
allowed $\mu_b =m$ such asymmetry could not be realized, then bosonic
condensate must be formed and the equilibrium distribution becomes:
\be
f_{b} = (2\pi)^3 m^3 C\delta({\bf p})+
\frac{1}{\exp\left(\frac{E_b-m}{T}\right)-1}
\label{f-bc}
\ee
where ${\bf p}$ is the three-momentum of the bosons. Thus in the case of
a large charge asymmetry the equilibrium plasma is still described by two
parameters/functions, $T(t)$ and $C(t)$, the latter being the number density
of particles in the condensate in units $m^3$.

The magnitude of chemical potentials of massless fermions is determined 
by the conditions of thermal equilibrium and in the particular case under
consideration they are:
\be
\mu_q = \mu_l = \mu_\chi/4
\label{mu-ferm}
\ee
Here we have used the equilibrium condition
\be
\sum_{in} \mu_{in} = \sum_{fin} \mu_{fin},
\label{mu-equil}
\ee
where the summation is made over all particles in initial (``in'') and 
final (``fin'')
states participating in all relevant reactions. For the system with two
conserved charges, $Q_B$ and $Q_L$ in the case under consideration, there
should be two independent chemical potentials. We excluded one of them
imposing the condition $Q_B-Q_L =0$. 

The energy and, say, lepton charge density of particle species $j$ are 
given respectively by:
\be 
\rho_j &=& \frac{1}{2\pi^2}\int dp p^2 E 
\left[f_j(E)+\bar f_j(E) \right] 
\label{rhoj}\\
p_j &=& \frac{1}{6\pi^2}\int dp p^2\,\frac{p^2}{ E}\, 
\left[f_j(E)+\bar f_j(E) \right] 
\label{pj}\\
L_{j} &=& \frac{Q_{Lj}}{2\pi^2}\int dp p^2 E 
\left[f_j(E)- \bar f_j(E) \right]
\label{Qj}
\ee
where $\bar f$ is the distribution function for antiparticles and
$L_j$ is the leptonic charge of particle $j$. In equilibrium
chemical potentials of antiparticles are opposite to those of particles, 
$\bar\mu = -\mu$.
We assume that $Q_{L_l}=Q_{L_\chi} = 1$.
If chemical potential of $\chi$ is positive and equal to $m$, 
then $C\neq 0$ for $\chi$, 
while $\bar C = 0$ for anti-$\chi$, and vice versa if $\mu_\chi=-m$.
Integration over angles in eqs. (\ref{rhoj}-\ref{Qj}) is trivially performed
because the distribution functions in homogeneous and isotropic universe
are also isotropic.

For massless quarks and leptons the integrals can be taken analytically.
The total energy density of all relativistic fermions plus photons
is equal to:
\be
\rho_{rel} = \rho_f +\rho_\gamma 
= {N_f T^4}\,\left( \frac{7\pi^2}{120} + \frac{\xi^2_f}{4}+
\frac{\xi^4_f}{8\pi^2} \right) +\frac{\pi^2}{15}\,T^4
= m^4\left(\gamma_1+\gamma_2\xi^{-2} +\gamma_3\xi^{-4}\right),
\label{rho-rel}
\ee
where $\xi_f =\mu_f/T = \mu_\chi/(4T)$ is dimensionless chemical potential
of fermions, $\xi = \mu_\chi/T$ is the chemical potential of $\chi$, and 
\be
\gamma_1=\frac{N_f}{2^{11}\pi^2},\,\,\, \gamma_2=\frac{N_f}{2^6},\,\,\, 
\gamma_3=\frac{\pi^2}{15}\left( \frac{7N_f}{8}+1\right).
\label{gammas}
\ee
Here $N_f$ is the total number of massless fermion species, quarks and
leptons included. For three quark-lepton families, three quark colors, and two
spin states: $N_f =48$ (the contribution of antiparticles is already included
inside the brackets of eq. (\ref{rho-rel})).

The charge density of massless leptons is: 
\be
L= \frac{N_l T^3 \xi_l}{6}\,\left(1 + \frac{\xi^2_l}{\pi^2}\right)
= m^3\left(\beta_0 + \beta_1\xi^{-2} \right)
\label{L}
\ee
where $\xi_l = \mu_\chi/4T$, $N_l = 12$ is the number of lepton species 
which include 3 families of particles with two spin states, 
and
\be
\beta_0 = \frac{N_l}{6\cdot 2^6 \pi^2},\,\,\, 
\beta_1 = \frac{N_l}{24}.
\label{betas}
\ee

The subtle points are the number of right-handed neutrino states and the
contribution of heavy t-quarks which may be absent at relatively small T
created after $\chi$-evaporation.
In particular, if the number of neutrino species is twice smaller due to
an absence of $\nu_R$, then $N_l$ would be 9, and $N_f=45$; if instead we
consider 6 neutrinos but exclude the top quark we would find $N_l=12$
and $N_f=42$. In both these cases it can be easily
seen that the equilibrium of the reaction $\chi \rar 3q+l$, together with
the condition $Q_B=Q_L$, no longer leads us to the simple relation
$\mu_q=\mu_l=\mu_\chi/4$. Indeed we find different values of chemical
potentials of quarks and leptons in terms of that of the $\chi$ particles.
We have done calculations for these two different scenarios, with the
calculated values for $\mu_q$ and $\mu_l$, and we have found that the
picture is unchanged, and all the results presented below are 
practically the same for different particle content.

The integrals which determine the energy and charge density of 
massive $\chi$-field cannot be taken analytically but in the limit of
low temperature, $T\ll m$, their approximate expressions are easily
found. In what follows we express everything in units of $\chi$-mass, $m$. 
In particular, when condensate is non-vanishing, i.e. $\mu=m$ and 
$C\neq 0$ we obtain:
\be
m^{-4}\rho_\chi &\approx& {C} + \frac{T^4\xi^{5/2}}{\sqrt{2} \pi^2m^4}
\int \frac{dy \sqrt{y}}{e^y -1} 
\left( 1 + \frac{9}{4}\,\frac{y}{\xi}\right)
= C + \alpha_1 \xi^{-3/2} + \frac{9}{5}\,\alpha_2 \xi^{-5/2},
\label{rho-nr}\\
m^{-4}p_\chi &\approx& \frac{\sqrt{2}\,T^4\,\xi^{3/2}}{3\pi^2 m^4}
\int \frac{dy y^{3/2}}{e^y -1}
= \alpha_3\xi^{-5/2}, 
\label{p-nr}\\
m^{-3}Q_\chi &\approx& C + \frac{T^3 \xi^{3/2}}{\sqrt{2}\pi^2m^3} 
\int\frac{dy \sqrt{y}}{e^y -1} 
\left( 1 + \frac{5}{4}\,\frac{y}{\xi}\right)=
C + \alpha_1 \xi^{-3/2} + \alpha_2 \xi^{-5/2},
\label{l-nr}
\ee
where $\alpha_1 = 0.164$, $\alpha_2 = 0.16$, and $\alpha_3 = 8\alpha_2/15$.
Notice that the condensate creates zero pressure because of 
$\delta({\bf p})$ in the distribution function (\ref{f-bc}).

\section{Condensate decay and plasma temperature \label{s-decay}}

Since by assumption the total leptonic and baryonic charges were 
conserved when $\chi$ relaxed down to zero in $U(1)$-symmetric 
quadratic potential (\ref{U-of-chi}) and decayed into channel (\ref{chi-decay}), 
each of them satisfied the conservation equation:
\be
\dot Q_{tot} + 3H Q_{tot} = 0.
\label{dot-q}
\ee
Thus leptonic or baryonic charge density evolved as:
\be
Q_{tot} \sim Q_{in} /a^3
\label{Q-of-a}
\ee

We have two unknown functions $T(a)$ and $C(a)$ if condensate is present
or $T(a)$ and $\mu(a)$ after condensate evaporation. To determine the law
of their evolution we need another equation which is the law of covariant
conservation of the total energy-momentum tensor. In 
Friedman-Robertson-Walker (FRW) metric this conservation law has the form:
\be
\dot \rho_{tot} = - 3 H \left( \rho_{tot} + p_{tot}\right),
\label{dot-rho}
\ee
where $\rho_{tot}$ includes the energy density of relativistic particles
(photons and fermions), and the energy density of $\chi$, either relativistic 
or not, depending upon the initial conditions. The pressure of relativistic
matter is $p_{rel} = \rho_{rel}/3$ and the pressure of $\chi$ is given by
eq. (\ref{pj}) which in non-relativistic case turns into (\ref{p-nr}).

It is convenient to look for the evolution of $C$ and $T$ as functions of
the scale factor $a$, taking the latter as the independent variable. In this
case the Hubble parameter disappears from the equation and we obtain:
\be
\rho'_{tot} = -3 \left( \rho_{tot} + p_{tot}\right)/a,
\label{rho'}
\ee
where prime means derivative with respect to the scale factor $a$.

\begin{figure}
\includegraphics{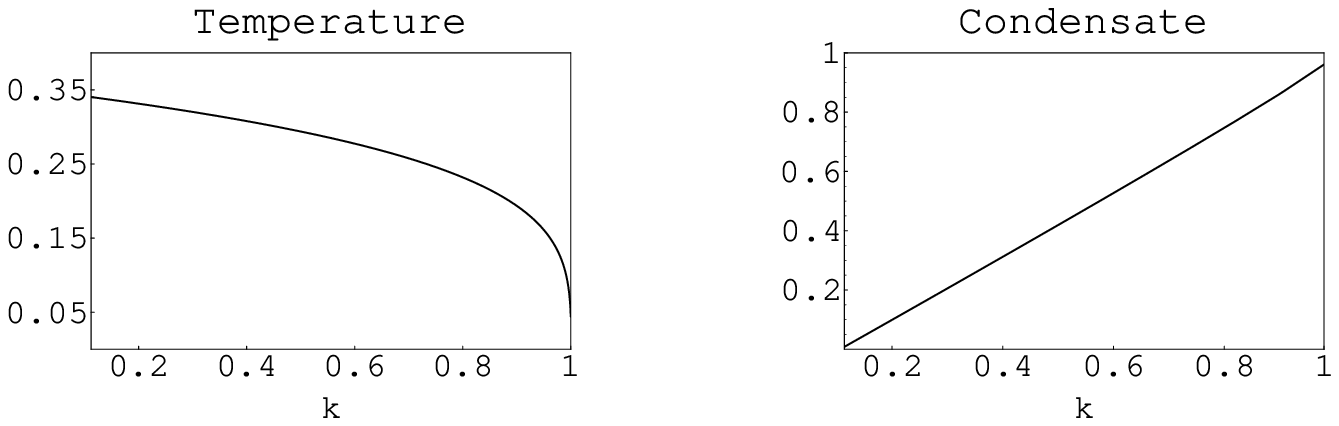}
\caption{\footnotesize{Initial values of temperature and condensate as functions of $\kappa$, with the fixed value of the total energy density $m^{-4}\rho_{tot}=1$.}}
\label{TCk-rho1}
\end{figure}

\begin{figure}
\includegraphics{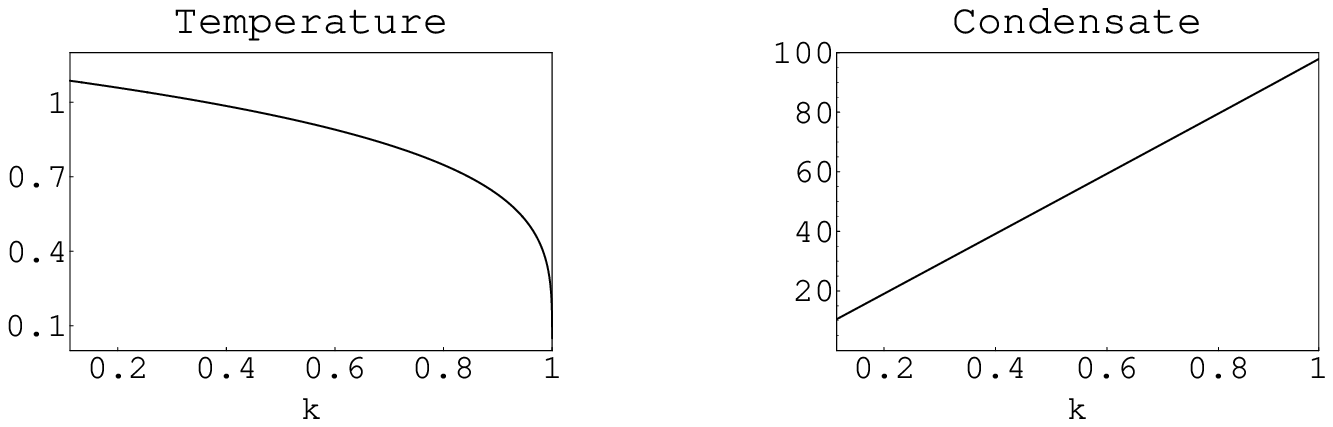}
\caption{\footnotesize{Initial values of temperature and condensate as functions of $\kappa$, with the fixed value of the total energy density $m^{-4}\rho_{tot}=100$.}}
\label{TCk-rho100}
\end{figure}

\begin{figure}
\includegraphics{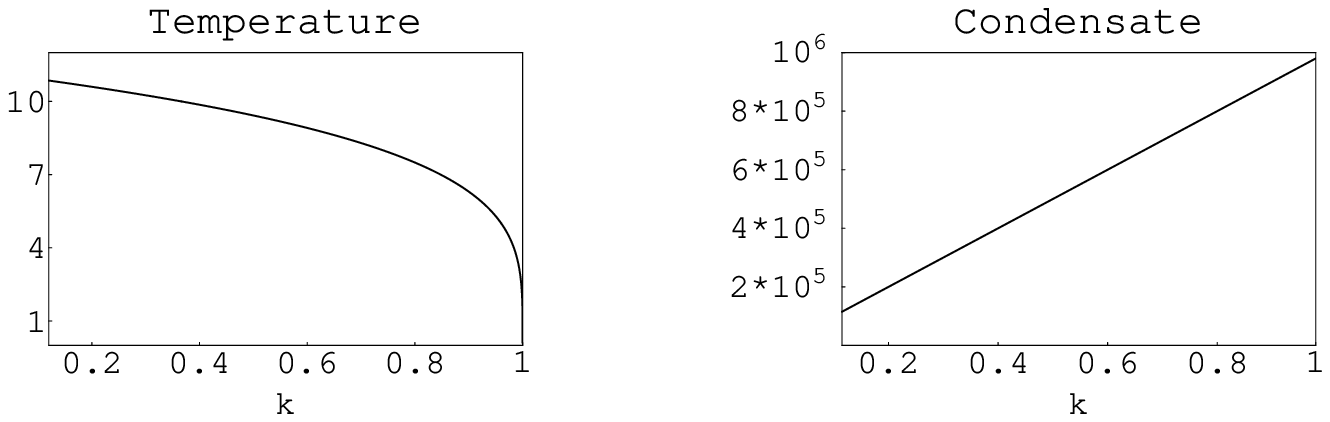}
\caption{\footnotesize{Initial values of temperature and condensate as functions of $\kappa$, with the fixed value of the total energy density $m^{-4}\rho_{tot}=10^6$.}}
\label{TCk-rho1000000}
\end{figure}

\begin{figure}
\includegraphics{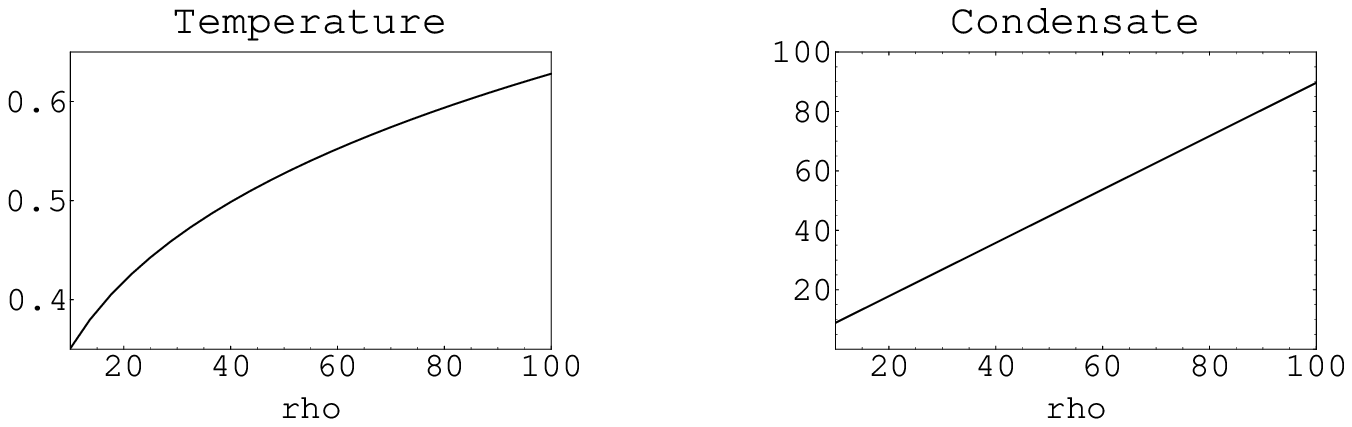}
\caption{\footnotesize{Initial values of temperature and condensate as functions of the energy density, with fixed $\kappa=0.9$.}}
\label{TCrho-k09}
\end{figure}

\begin{figure}
\includegraphics{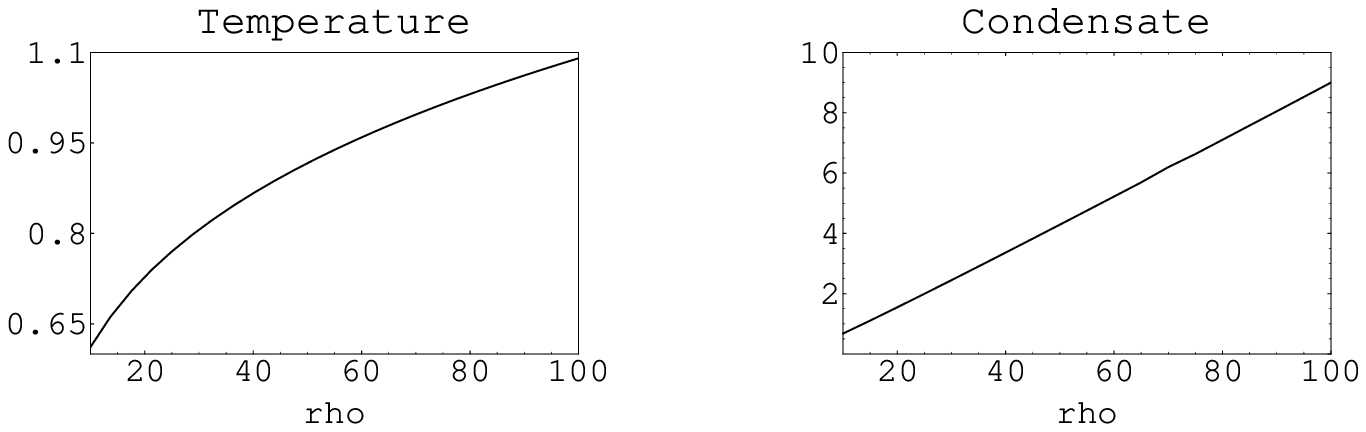}
\caption{\footnotesize{Initial values of temperature and condensate as functions of the energy density, with fixed $\kappa=0.1$.}}
\label{TCrho-k01}
\end{figure}

The initial values of $\xi = m/T$ and $C$ can be expressed through the
initial values of the total energy density, $\rho^{(in)}$ and, say, 
leptonic charge density, $Q_l^{(in)}=\kappa\rho^{(in)} $. As we have
already mentioned $|\kappa|\leq 1$ or it may be slightly larger than 1.
The values of $C^{(in)}$ and $T^{(in)}$
are presented in figures \ref{TCk-rho1} to \ref{TCrho-k01} 
as functions of $\rho^{(in)}$ and $\kappa$.
In particular, figs. \ref{TCk-rho1} to \ref{TCk-rho1000000} show how 
$T_{in}$ and $C_{in}$ depend upon $\kappa$ when $\rho_{in}$ is fixed, 
while figs. \ref{TCrho-k09} and \ref{TCrho-k01} deal with
their dependence upon the total $\rho_{in}$, with fixed $\kappa$.
If $\kappa\approx 1$ the initial temperature is low and 
$\xi^{(in)} \gg 1$. In particular, for $\kappa =1$ the initial temperature
does not depend upon the initial value of $\rho$ and is equal to:
\be
T_{in}\approx0.05 m
\label{T-in}
\ee
However, even if $\kappa$ is slightly smaller than 
unity, $T^{(in)}$ becomes much higher and, for large $\rho_{in}$, even relativistic 
initial state could be realized, i.e. $m/T <1$. E.g. for $\kappa = 0.9$ 
and $\rho_{in} = 10^3$ the initial temperature is $T_{in}\approx m$.
Such a large rise of temperature
is explained by much larger energy release into plasma by $\chi \bar\chi$
annihilation, which is absent in the case of $|\kappa| =1$ when antiparticles
of $\chi$ are initially absent.

We have numerically integrated this equation with 
constraint (\ref{Q-of-a}). The solution is divided into two parts:
initially with non-zero and running $C(a)$ (condensate), $\mu= m$, 
and running $T(a)$. Then at some
stage $C$ reaches zero and the condensate disappears. After that the
solution is searched for $C\equiv 0$ but running $\mu (a)$ and $T(a)$.
The details are presented in the Appendix. The calculations are very much simplified
in the limit of low temperatures when $T\ll m$. The functions $\rho_{tot}$
and $p_{tot}$ in this limit are simple algebraic functions and not integrals
over energy and the evolution equation turns into:
\be 
\frac{\xi'}{\xi}\, 
\frac{2\alpha_2\xi^{-5/2}+\beta_1\xi^{-2}+4\gamma_3\xi^{-4}}
{4\alpha_2\xi^{-5/2}+3\beta_1 \xi^{-2}+4\gamma_3\xi^{-4}}=\frac{1}{a}
\label{xi'}
\ee

Depending upon which term dominates in the $\xi$ dependent coefficients
in the l.h.s. of this equation we obtain different regimes of the 
cosmological cooling. At relatively small $\xi$, when $\xi^{-4}$ dominates,
the temperature drops as $T\sim 1/a$. It is the usual law of relativistic 
cooling. If the first term, $\xi^{-5/2}$, dominates, $T\sim 1/a^2$. This is
also the known regime for dominating non-relativistic matter. However, if
the term $\xi^{-2}$ is dominant, the temperature drops unusually fast:
\be
T\sim 1/a^3
\label{Ta3}
\ee
We have found that this regime is the typical one in the course of the 
condensate evaporation, so the temperature becomes very low at the end.
It is quite unusual situation. Normally one would expect that a source
of energy created by the evaporating condensate would lead to plasma
heating and a slower decrease of the temperature in the course of expansion.
We observe an opposite picture. Probably this phenomenon is induced by a very
slow decay of the condensate and efficient cooling of the particles in
the plasma by the scattering on the ``refrigerator'' created by a large 
number of cold $\chi$-particles at rest.

Here we would like to remark that the derivation of the temperature's 
evolution law can be extended to the case of a different particle content, 
even if eq. (\ref{mu-ferm}) is no longer true, as we have already pointed
out. In fact in such a system, when both energy and charge densities are
dominated by $\xi^{-2}$ terms, we can generalize eq. (\ref{Ta3}) to the following:
\be
T\sim 1/a^{3z}
\label{Ta3z}
\ee
where:
\be
z=\frac{2N_l\mu_l^2+2N_f\mu_f^2-N_l\mu_l}{3N_l\mu_l^2+3N_f\mu_f^2-2N_l\mu_l}
\label{z}
\ee
If we now insert the chemical potentials given by the equations of equilibrium
with respect to reaction (\ref{chi-decay}), together with the equality of
charges $Q_B-Q_L =0$, we will find for a different particle contents, that $z=1$,
which brings back eq. (\ref{Ta3}).

In non-relativistic case, as well as in the relativistic one, with
dominating condensate, the behavior of $C(a)$ can be predicted analytically
from evolution equations (\ref{dot-q}) and (\ref{dot-rho}):
\be
C' = -\frac{3}{a}\left(C-u(T)\right)
\label{C'}
\ee
where $u(T)$ contains the temperature-dependent terms which contribute to
the energy or to the charge density. Thus if the condensate dominates 
the energy and charge densities, i.e. $C\gg u(T)$, the evaporation is
described by the law $C\sim 1/a^3$.

The evolution of the interesting quantities is presented in 
figures \ref{Evol1} and \ref{Evol2}. In accordance with the arguments
presented above the amplitude of the condensate follows the simple law:
\be
C\sim 1/a^3
\label{Ca3}
\ee
which holds through most of its life. Only when the energy density 
of the condensate becomes approximately equal to
that of the plasma, $C(t)$ starts to drop faster and at some moment
it reaches zero. All the plots with $\mu = m=const$ have been stopped at 
the precise moment of the complete evaporation.

If the initial temperature is high, we do not have an accurate analytical
approximation for the energy, pressure, and charge density of $\chi$ with 
the large chemical potential $\mu=m$. In this case we need to solve differential
equation (\ref{rho'}) with the integrals for $\rho$, $p$, and $Q$ taken 
numerically. The solutions are presented in fig. \ref{Evol-rel}. 
Initially the temperature drops as $T\sim 1/a$ and later, when it becomes 
smaller than $m$ before the complete evaporation of the condensate, the
regime changes to that discussed above in the non-relativistic case.

\section{Results and discussion\label{s-results}}

\begin{figure}
\begin{center}
\includegraphics{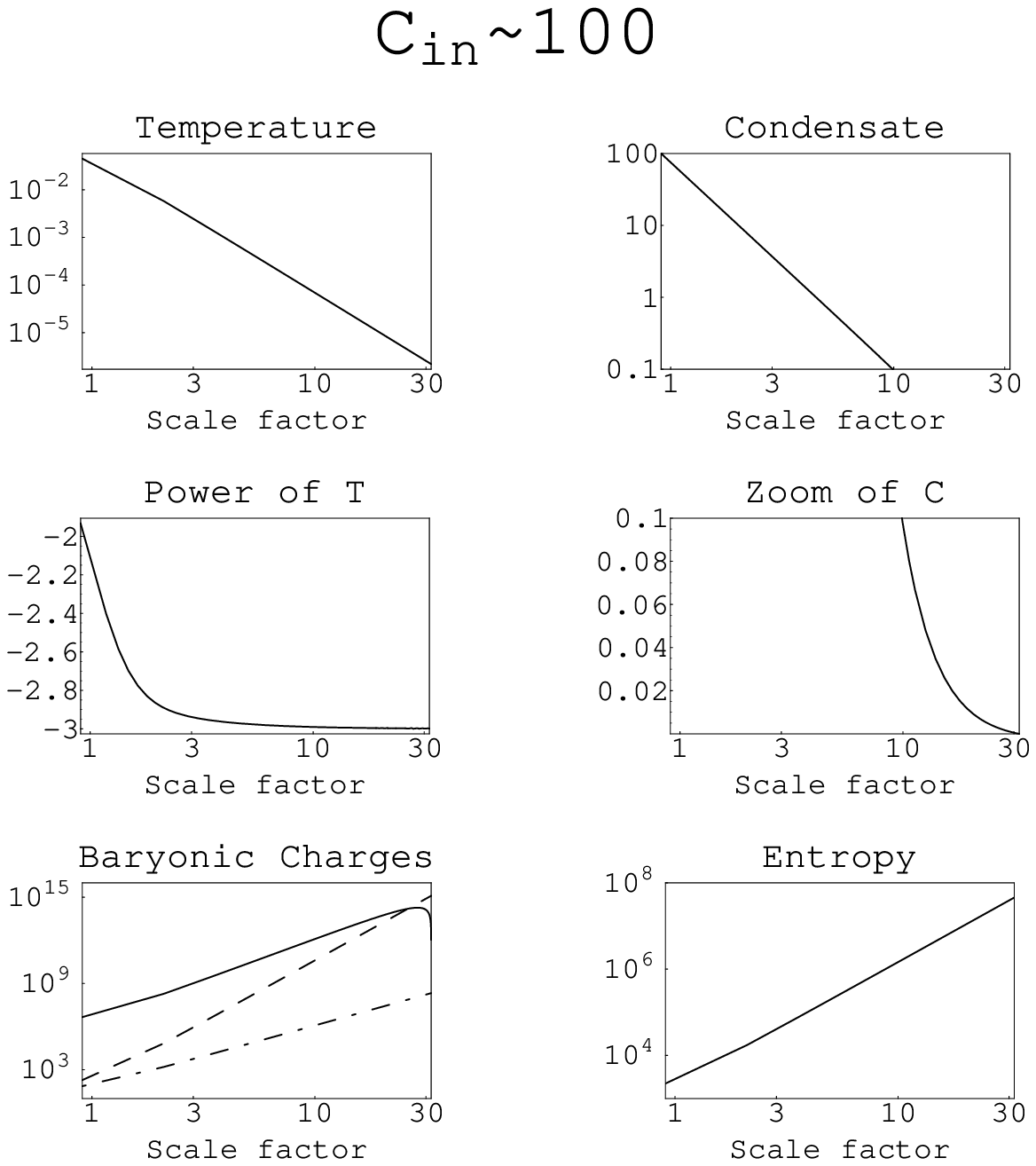}
\caption{\footnotesize{This figure refers to the special case $\kappa=1$. 
Clockwise from upper left corner: evolution of temperature, evolution of 
the condensate (upper right), the complete evaporation of the condensate
(middle right), evolution of entropy in comoving volume (lower right),
evolution of baryonic charges (lower left corner), and the power of $T(a)$
(middle left), as functions of the scale factor $a$. The 'Baryonic Charges' graph has 
the following lines: solid line is for the condensate; dashed line is for $\chi$ 
and $\bar{\chi}$ particles; dot-dashed line is for relativistic charged particles.}}
\label{Evol1}
\end{center}
\end{figure}

\begin{figure}
\begin{center}
\includegraphics{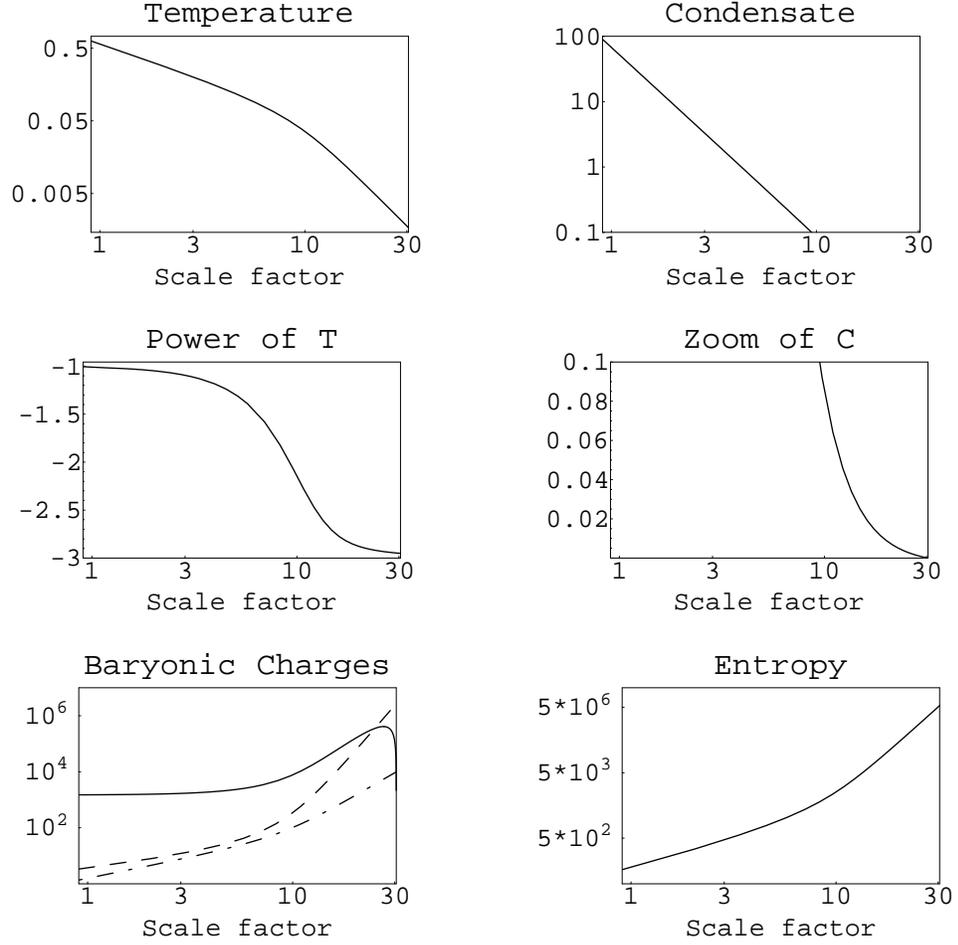}
\caption{\footnotesize{This figure refers to the case $\kappa=0.9$, 
while $m^{-4}\rho_{tot}=100$. 
Clockwise from upper left corner: evolution of temperature, evolution of 
the condensate (upper right), the complete evaporation of the condensate
(middle right), evolution of entropy in comoving volume (lower right),
evolution of baryonic charges (lower left corner), and the power of $T(a)$
(middle left), as functions of the scale factor $a$. The 'Baryonic Charges' graph has 
the following lines: solid line is for the condensate; dashed line is for $\chi$ 
and $\bar{\chi}$ particles; dot-dashed line is for relativistic charged particles.}}
\label{Evol2}
\end{center}
\end{figure}

\begin{figure}
\includegraphics{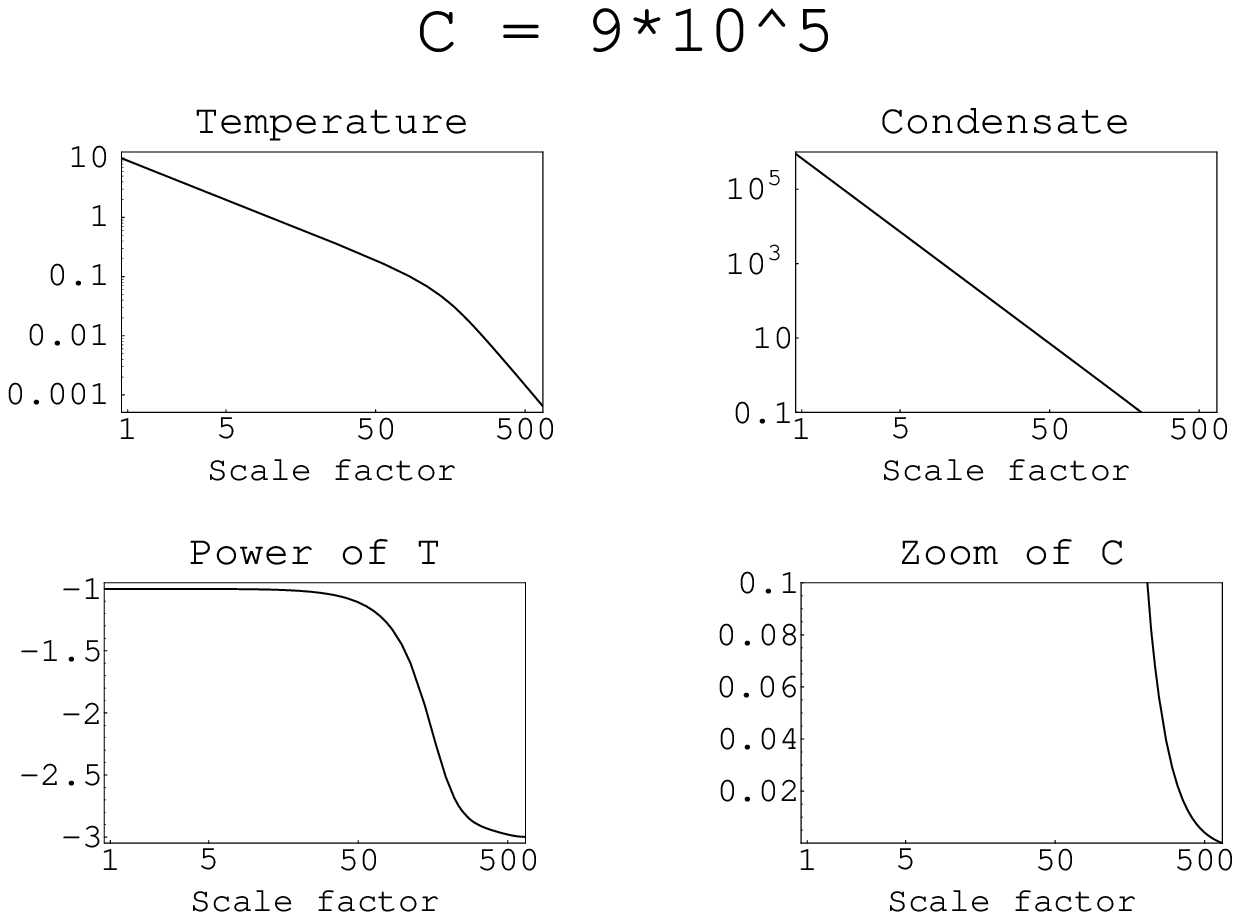}
\caption{\footnotesize{This figure refers to the relativistic initial state, 
that is $T>m$. Clockwise from upper left corner: evolution of temperature, 
evolution of the condensate, the complete evaporation of the condensate, 
the power of $T(a)$, as functions of the scale factor $a$. We see that the 
evaporation happens only when temperature has dropped below $m$.}}
\label{Evol-rel}
\end{figure}

\begin{figure}
\includegraphics{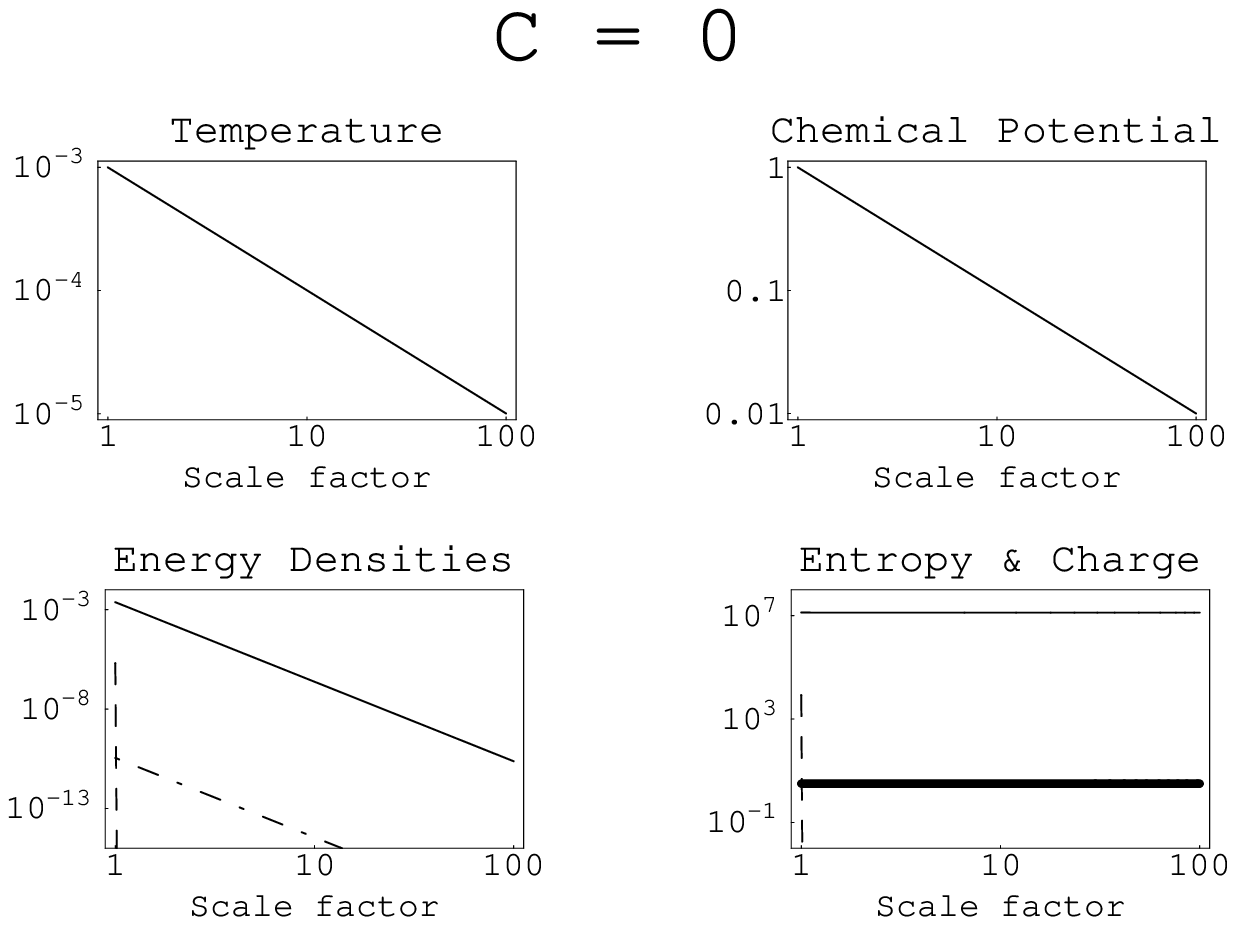}
\caption{\footnotesize{This figure refers to the second part of the problem, 
when the condensate has completely evaporated and $\mu$ is no longer a constant. 
Clockwise from upper left corner: evolution of temperature, chemical potential, 
entropy in comoving volume and baryonic charges, and energy densities, as 
functions of the scale factor $a$. The 'Energy densities' graph has the following 
lines: solid line is for relativistic charged particles; dashed line is for 
photons; dot-dashed line is for $\chi$ and $\bar{\chi}$ 
particles. The 'Entropy \& Charge' graph has the following lines: 
the solid line is 
the charge of relativistic particles; the dashed line is the charge of $\chi$ and 
$\bar{\chi}$ particles, and the thick solid line is the entropy.}}
\label{Evol-zeroC}
\end{figure}

The obtained results are presented in figs. \ref{Evol1}, \ref{Evol2},
\ref{Evol-rel}, and \ref{Evol-zeroC}. All the quantities which appears in
these figures are normalized to a proper power of the mass of $\chi$, namely:
\begin{eqnarray}
	T\rar T/m, & \mu\rar \mu/m, \nonumber\\
	Q\rar Q/m^3, & \rho\rar \rho/m^4 \nonumber
\end{eqnarray}
and they are expressed as functions of the scale factor ratio $a/a_{in}$.
Most of these graphs are in the log-log scale, except for the `Power of T'
and `Zoom of C' plots, which are in the log-linear scale.

In fig. \ref{Evol1} a special case $\kappa=1$ is presented.
We see in the upper left box how the temperature evolves,
while the box below (middle left), named `Power of T',
shows that the cooling follows the law $T\sim1/a^3$.
The same law holds for the condensate, as it is shown in the right upper
box, that is, $C\sim1/a^3$, while the middle right box called
`Zoom of C', is a zoom of its final evolution, and shows the instant of
the complete evaporation. The calculated cosmological charge asymmetry,
$\beta = Q/n_{\gamma} = Q/0.24\,T^3$,
is presented in the left lower box: there are lines for the condensate
(solid line), decay products (dashed line), and relativistic light charged
particles (dot-dashed line).
It can be seen that the produced asymmetry strongly rises during
the non-relativistic regime.
This is understandable because the temperature drops very fast,
$T^3\sim a^{-9}$ while the charge density drops only as $Q\sim a^{-3}$. We have
found that the ratio of $\beta(a=1)$ to $\beta$ at $C=0$ is, in this
case, of the order of $10^8$. The last box (lower right) shows how entropy in
comoving volume, $S = s a^3 = a^3(\rho+p)/T$, evolves: we see, as it
was for $\beta$, that the fast cooling results in $S$ rising as $a^3$
because the main contribution to the energy density comes from the
condensate, which drops only as $1/a^3$.

In fig. \ref{Evol2} the same graphs are presented, but for
different initial values of temperature and condensate which 
correspond to $\kappa=0.9$, and respectively to $T_{in}\approx 0.65m$.
We see that all the quantities evolve similarly to the previous case
presented in figure \ref{Evol1}, but here one
can better see how the behavior of temperature
switches from $T\sim1/a$ to $T\sim1/a^3$ with growing scale factor.
The moment when it happens is precisely the moment when the 
$\xi^{-2}$-terms in the energy and charge densities start 
to dominate over the standard $\xi^{-4}$ ones, see eq. (\ref{xi'}). 
In this situation we have
found the ratio of the final to initial charge density to be about
$10^3$.

Next, in fig. \ref{Evol-rel} the case of initial relativistic
state is presented. In order to have relativistic initial state we
had to choose $\kappa=0.1$ and a huge initial total energy density,
$\rho_{in}\approx 10^6m^4$. An important thing to be noticed here is that
the condensate can evaporate completely only when the non-relativistic
limit is reached, i.e. $T< m$.

Finally, in fig. \ref{Evol-zeroC} we present evolution of temperature,
chemical potential, energy densities, baryonic charge, and entropy, when the
condensate has completely evaporated. In the first box, clockwise from upper
left corner, we see that the temperature follows the standard evolution law
for a relativistic plasma, $T\sim1/a$. In the second box we have the same
graph for the chemical potential, which also behaves as $\mu\sim1/a$. The
third box contains: entropy in comoving volume (thick solid line); charge
density, or to be more precise $\beta$, contributed by relativistic 
(thin solid line)
and non-relativistic (thin dashed line\footnote{This line, as 
well as the thin dashed
one in the fourth box, seems to be vertical but in fact it is not: it is an
extremely fast decrease of the contribution of 
non-relativistic $\chi$ and $\bar\chi$ particles.}) 
species. Finally, the fourth box (lower left corner) shows the
evolution and the ratios of the energy densities of charged relativistic
particles (solid line), photons (dot-dashed line), and non-relativistic
$\chi$-particles (dashed line). We see that both entropy in
comoving volume and $\beta$ are conserved. One may wonder if a non-zero 
chemical potential could break the entropy conservation law, but it
can be easily checked that this is not so because
every term in the expression of charge density and
entropy in comoving volume behave as $1/a^3$.
This, together with the fact that
the energy density of non-relativistic particles (as well as their charge density)
is orders of magnitude smaller than that of the relativistic ones, as one can
see in the 4th box, gives the result
just obtained. Hence the Affleck and Dine scenario of baryogenesis could
indeed lead to a huge baryon asymmetry of the universe.

To dilute the asymmetry to the observed level a subsequent
release of entropy is necessary or evaporation into initially hot
plasma. Both such mechanisms were discussed recently in 
ref.~\cite{dksy}, where one can find references to many earlier papers. 

\appendix

\section{Appendix}
\paragraph{}
In this appendix we present some details of the
numerical solution of the differential 
equations\footnote{All the quantities here are normalized to
proper powers of $\chi$-mass, e.g. $T\rar T/m$, $\rho\rar\rho/m^4$.}:
\be
\left\{
\begin{array}{ll}
\dot Q_{tot} = - 3H Q_{tot} \\
\dot \rho_{tot} = - 3 H \left( \rho_{tot} + p_{tot}\right) 
\end{array}
    \right. 
\begin{array}{ll}
\Rightarrow
\end{array}
\left\{
\begin{array}{ll}
Q'_{tot} = -3\;Q_{tot}/a \\
\rho'_{tot} = -3 \left( \rho_{tot} + p_{tot}\right)/a
\end{array}
    \right.
\label{q-and-rho-together}
\ee
We have to insert into these expressions the explicit values of $Q_{tot}$
and $\rho_{tot}$, which depend either upon $T(a)$ and $C(a)$ when $\mu=m$, 
or upon $T(a)$ and $\mu(a)$ when $C=0$, that is:
\be
\left\{
\begin{array}{ll}
Q_{tot} = Q_{tot}(T,C) \\
\rho_{tot} = \rho_{tot}(T,C)
\end{array}
    \right. 
\begin{array}{ll}
\textrm{or} 
\end{array}
\left\{
\begin{array}{ll}
Q_{tot} = Q_{tot}(T,\mu) \\
\rho_{tot} = \rho_{tot}(T,\mu)
\end{array}
    \right.
\label{q-and-rho-dependance}
\ee
In the first case, when $C\neq0$, these equations read:
\begin{equation}
\left\{
\begin{array}{ll}
                    Q_{tot}' = C' + \zeta\; T' \\
                    \rho'_{tot} = C' + \eta\; T'
                \end{array}
    \right.
\end{equation}
where $\zeta(T)$ and $\eta(T)$ are the derivatives
of $Q_{tot}(T,C)$ and $\rho_{tot}(T,C)$ with respect to $T$.
The system can be solved together with 
equations (\ref{q-and-rho-together}) to have:
\begin{equation}
    \left\{
  \begin{array}{ll}
(\eta-\zeta)\;T' =  3\left[Q_{tot}-(\rho_{tot}+p_{tot})\right]/a \\
                    \\
 (\eta-\zeta)\;C' =  3\left[\zeta(\rho_{tot}+p_{tot})-\eta\; 
Q_{tot}\right]/a
                \end{array}
    \right.
\label{system1}
\end{equation}
As for the second case, when $C=0$, the derivatives of $Q_{tot}$ and 
$\rho_{tot}$ are:
\begin{equation}
    \left\{
                \begin{array}{ll}
                    Q_{tot}' = f\;\mu' + g\;T' \\
                    \rho'_{tot} = h\;\mu' + l\;T'
                \end{array}
    \right.
\end{equation}
where $f(\mu,T)$ and $h(\mu,T)$ are the derivatives of
$Q_{tot}(T,\mu)$ and $\rho_{tot}(T,\mu)$ with respect
to $\mu$, while $g(\mu,T)$ and $f(\mu,T)$ are the derivatives
of $Q_{tot}(T,\mu)$ and $\rho_{tot}(T,\mu)$ with respect to $T$.
Again, the system can be solved together with 
equations (\ref{q-and-rho-together}) to have:
\begin{equation}
    \left\{
  \begin{array}{ll}
 (gh-fl)\;T' =  3\left[f(\rho_{tot}+p_{tot})-h\;Q_{tot}\right]/a \\
                    \\
f(gh-fl)\;\mu' =  3\left[fl\;Q_{tot}-gf(\rho_{tot}+p_{tot})\right]/a
                \end{array}
    \right.
\label{system2}
\end{equation}
Now the system of differential equations (\ref{system2}), as well as system
(\ref{system1}), can be solved numerically.

\newpage
\clearpage

\end{document}